\documentstyle[12pt]{article}
\def\a{\begin{eqnarray}}
\def\b{\end{eqnarray}}
\def\0{\nonumber}

\begin{document}
\begin{flushright}
SISSA-ISAS 161/92/EP
\end{flushright}
\vskip0.5cm
\centerline{\LARGE\bf Matrix models without scaling limit}
\vskip2.3cm
\centerline{\large  L.Bonora}
\centerline{International School for Advanced Studies (SISSA/ISAS)}
\centerline{Via Beirut 2, 34014 Trieste, Italy}
\centerline{INFN, Sezione di Trieste.  }
\vskip0.5cm
\centerline{\large C.S.Xiong}
\centerline{International School for Advanced Studies (SISSA/ISAS)}
\centerline{Via Beirut 2, 34014 Trieste, Italy}
\vskip5cm
\abstract{In the context of hermitean one--matrix models
we show that the emergence of the NLS hierarchy and of its reduction,
the KdV hierarchy, is an exact result of the lattice characterizing
the matrix model. Said otherwise, we are not obliged to take a
continuum limit to find these hierarchies. We interpret this result
as an indication of the topological nature of them.
We discuss the topological field theories associated with both
and discuss the connection with topological field theories coupled to
topological gravity already studied in the literature.}

\vfill
\eject

\section{Introduction}

One--matrix models give rise to a lattice
hierarchy of (differential--difference) equations, as well as to
a discrete string equation. In the usual approach \cite{MM},
\cite{MMR}, one takes
a continuum limit and recovers (in the even potential
case) the KdV hierarchy and the continuum string equation. These two
objects can be interpreted both in terms of a topological field theory
and in terms of intersection theory of moduli space
of Riemann surfaces \cite{W1},\cite{W2},\cite{DW},\cite{Ko},\cite{D}.

In the present paper we show that both the KdV hierarchy and the
corresponding string equation are intrinsic to the matrix model
lattice---they are not
simply a result of an accidentally lucky scaling limit of the lattice,
as oral tradition seems sometimes to put it.
We do it by extracting the KdV hierarchy and string equations
directly from the lattice counterparts without passing through a limiting
procedure \footnote{However, since in the continuum limit we may have to
renormalize and relabel our objects, the two approaches are expected
to differ by renormalization or relabeling of the coupling constants.}.
The method consists essentially in using the first flow equation to eliminate
the difference operations in the remaining equations. The latter become
a hierarchy of differential equations.

This result is not surprising after all. Let us clarify our point of view
with the help of an analogy:
the Euler number of a Riemann surface $\Sigma$ of genus $h$ can be
calculated with a continuum procedure by means of
\a
2-2h= {1\over {4\pi}}\int_{\Sigma} \sqrt{g} R\label{c}
\b
where $R$ is the scalar curvature w.r.t. the metric $g$; but it can also be
calculated by means of
\a
2-2h= V-L+F \label{d}
\b
with reference to a simplicial approximation of $\Sigma$, in which
$V, L, F$ are the number of vertices, links and faces, respectively.

The KdV hierarchy and string equation obtained via a lattice scaling limit
are akin to the RHS of eq.(\ref{c}). The KdV hierarchy and string equation
we discuss in this paper are parallel to the RHS of eq.(\ref{d}).
Since KdV hierarchy and corresponding string equation have a topological
significance, the analogy just presented is more than a simple suggestion.
The same can be

Actually in this paper we do more. We study the complete one--matrix model
(not simply the even potential case), we define the corresponding lattice
hierarchy (the Toda lattice hierarchy), and, with the method of \cite{BX},
we map it to a differential hierarchy, the non linear Schr\"odinger (NLS)
hierarchy. The above mentioned KdV hierarchy is nothing
but a reduction of this hierarchy. Finally we discuss the topological field
theory associated to NLS. We show that
it belongs to the family of topological conformal models with two primaries
coupled to topological gravity studied in \cite{W1},\cite{DW}.

The paper is organized as follows. In section 2 we review  the method
presented in \cite{BX} to extract differential
hierarchies from the Toda lattice hierarchy. In section 3
we review and complete the reduction to the NLS hierarchy
and the corresponding string equation,
which corresponds to the most general one--matrix model.
In section 4 we examine the even potential case, i.e.
we show how to extract the KdV hierarchy from the NLS hierarchy and
the corresponding string equation. We also
discuss the corresponding well--known topological field theory.
In section 5 we propose a topological field theory corresponding to
the most general one--matrix model. It turns out to correspond
to the topological theory mentioned above.

\section{Review: from Toda lattice to differential hierarchies}

This section is a review of \cite{BX}. First some notations.
Given a matrix $M$, we will denote by $M_-$ the strictly lower triangular
part and by $M_+$ the upper triangular part including the main diagonal.
We deal with $\infty \times\infty$ matrices. In due time we will pass to
semi--infinite matrices. As usual $E_{ij}$ denotes the matrix
$(E_{ij})_{kl}=\delta_{ik}\delta_{jl}$.
We also use
\a
I_{\pm}\equiv \sum_{i=-\infty}^{\infty}E_{i,i\pm 1},
\quad\qquad
\rho=\sum_{i=-\infty}^{\infty}i E_{ii}\0
\b
Throughout this paper $\lambda$ denotes the spectral parameter,
and $\Lambda$ represents an infinite dimensional column vector with
components $\Lambda_n=\lambda^n,~n\in {\bf Z}$.
The vector $\Lambda$ is our elementary starting point. From it we obtain
$\eta$
\a
\eta=exp\bigl({\sum_{r=1}^{\infty}t_r\lambda^r}\bigl)\Lambda \label{eta}
\b
In (\ref{eta}) $t_r$ are time or flow parameters.
On $\eta$ one can naturally define a (elementary) linear system
\a
&&\lambda\eta=\partial\eta
=I_+\eta\0\\
&&\lambda^r\eta={\partial\over{\partial t_r}}\eta
=\partial^r\eta
=I^r_+\eta\label{etasystem}\\
&&{\partial\over{\partial \lambda}}\eta=P_0\eta,\qquad
P_0=\rho I_-+\sum_{r=1}^{\infty}rt_rI_+^{r-1}
\0
\b
Throughout the paper $\partial$ denotes the derivative ${\partial \over
{\partial  t_1}}$. It is easy to see that the spectral and flow equations
are automatically compatible.

Next we introduce the ``wave matrix" $W$:
\a
W= 1+\sum_{i=1}^{\infty}
\sum_{n=-\infty}^{+\infty}w_i(n)E_{n,n-i}  \label{W}
\b
where $w_i(n)$ are functions of the time parameters.  We impose them to be
determined by the equations of motion
\a
{\partial\over{\partial t_r}}W=Q^r_+W-WI_+^r, \quad\quad
Q=WI_+W^{-1}
\label{emW}
\b
Finally we introduce the vector $\Psi=W\eta$.
In terms of all these objects the dynamical system we have defined can be
written as
\a
Q\Psi&=&\lambda\Psi\0\\
{\partial\over{\partial t_r}}\Psi&=&Q^r_+\Psi\label{dLS}\\
{\partial\over{\partial \lambda}}\Psi&=&P\Psi, \quad\quad P= WP_0W^{-1}
\0
\b

The compatibility conditions of this linear
system form
the so--called {\it discrete} KP--hierarchy (Toda lattice hierarchy)
\a
{\partial\over{\partial t_r}}Q=[Q^r_+, Q]\label{dKP}
\b
together with the {\it trivial} relation
\a
[Q, P]=1\label{dstring}
\b

We have
\a
Q=I_++\sum_{i=0}^{\infty}\sum_{n=-\infty}^{+\infty}a_i(n)E_{n,n-i},
\quad\quad\quad
P=\sum_{r=1}^{\infty}rt_rQ^{r-1}+\sum_{i=0}^{\infty}v_iQ^{-i-1}\label{PQ}
\b
where $v_i$ are diagonal matrices. The $a_i(n)$ and $w_i(n)$ are different
coordinates of the system.

The Toda lattice hierarchy (\ref{dKP})
consists of an infinite set of differential-difference equations.
In the rest of this section we review the procedure to pass from the above
matrix formalism to a related (pseudo--differential) operator formalism and
obtain a new hierarchy which consists merely of differential equations.
This procedure is referred to above as the mapping from the discrete
hierarchy to the corresponding differentiable hierarchy.

The essential ingredient is the remark that equation (\ref{eta}) implies
\a
\eta_n=\partial^{n-m}\eta_m,\qquad\qquad \forall n,m:\hbox{\rm
integers}\0
\b
This leads to
\a
\Psi_n=(W\eta)_n=
\sum_{i=-\infty}^nW_{ni}\eta_i
=\sum_{i=-\infty}^nW_{ni}\partial^{i-n}\eta_n
={\hat W}_n\eta_n
\label{Psin}
\b
where
\a
{\hat W}_n=1+\sum_{i<n}W_{ni}\partial^{i-n}
=1+\sum_{i=1}^{\infty}w_i(n)\partial^{-i}
\label{Wn}
\b
i.e. the ``wave matrix" $W$ can be considered as an infinite diagonal
matrix, whose components are invertible differential operators.
Using this fact we can transform the linear system (\ref{dLS}) into the
differential linear system
\a
&&{\hat Q}_n\Psi_n=\lambda\Psi_n\0\\
&&{\partial\over{\partial t_r}}\Psi_n
=\bigl({\hat Q}_n^r\bigl)_+\Psi_n, \qquad \forall n~{\rm integer}
\label{cLS}\\
&&{\partial\over{\partial\lambda}}\Psi_n=\bigl(\sum_{r=1}^{\infty}rt_r
{\hat Q}_n^{r-1}+\sum_{i=0}^{\infty}v_i(n)
{\hat Q}_n^{-i-1}\bigl)\Psi_n\equiv \hat P_n \Psi_n\0
\b
where the subscript ``+" selects now the non--negative
powers of the derivative $\partial$, and
\a
{\hat Q}_n={\hat W}_n\partial {\hat W}_n^{-1}
=\partial+\sum_{i=1}^{\infty}u_i(n)\partial^{-i}\qquad\forall
n~ {\rm integer}\label{Qn}
\b
The compatibility conditions are
\a
{\partial\over{\partial t_r}}{\hat Q}_n=[\bigl({\hat Q}_n\bigl)^r_+,
{\hat  Q}_n]\label{cKP}
\b
beside the trivial one
\a
\relax [\hat Q_n, \hat P_n]=1\label{cstring}
\b
Eq.(\ref{cKP}) specifies the {\it differential} hierarchy promised above.
Actually we have an infinite number of hierarchies, one for each integer
$n$. However they are not independent as they are related by the $t_1$
flow; we will see below that in the particular case of one-matrix models
all these hierarchies are isomorphic.

\section{Reductions corresponding to one-matrix models. The NLS hierarchy.}

It is well known \cite{BMX} that the data of a one--matrix model  can be
encoded in a corresponding discrete linear system.
Here we refer to hermitean one--matrix models, specified by an $N\times N$
hermitean matrix $M$, with partition function and potential given by
\a
Z_N(t)= \int dM e^{Tr V(M)},\qquad V(M)= \sum_{r=1}^\infty t_r M^r\0
\b
Henceforth we will consider reductions of the large system introduced
in the previous section so that they reproduce systems corresponding
to these one--matrix models.
Our purpose is to obtain the three crucial ingredients of matrix
models: the identification of the partition function, the hierarchy of
differential equations that give us the flows with respect to the coupling
parameters and the string equation.

First of all we reduce the system to semi--infinite matrices. In other
words the index $n$ in eqs.(\ref{W},\ref{PQ}) will run from now on
from 0 up to $+\infty$. With this proviso
the case relevant to one--matrix models is specified by the conditions
\a
a_0(n)=S_n,\qquad
a_1(n)=R_n,\qquad
a_i(n)=0,\qquad\forall n\geq2\0
\b
The discrete $t_1$ flow gives
\a
&&{\partial\over {\partial t_1}} \psi_n= \psi_{n+1} +S_n \psi_n
\label{psin}\\
&&{\partial\over{\partial t_1}}S_n=R_{n+1}-R_n\label{Sn}\\
&&{\partial\over{\partial t_1}}R_n=R_n(S_n-S_{n-1})\label{Rn}
\b
while the spectral equation (see (\ref{dLS})) writes
\a
\lambda \psi_n = \psi_{n+1} +S_n \psi_n + R_n \psi_{n-1}=
\partial \psi_n + R_n \frac {1} {\partial -S_{n-1}}\psi_n\0
\b
due to (\ref{psin}).

Therefore, in this case, the pseudo--differential
operator $\hat Q_n$ is
\a
\hat Q_n = \partial +R_n \frac {1}{\partial- S_{n-1}}\0
\b
Now (\ref{Rn}) gives
\a
S_{n-1}=S_n-{{R_n^{'}}\over{R_n}}\0
\b
and it is easy to prove that
\a
&&{\hat Q}_n =\partial+\sum_{l=1}u_l(n)\partial^{-l}\label{hQn}\\
&&u_l(n) = (-\partial + S_n)^{l-1}R_n \label{ul}
\b
In these equations a prime denotes derivative with respect to $t_1$.
As anticipated in the previous section the $u_l(n)$'s depend only on the
coordinates $S_n$ and $R_n$ and their derivatives. As a consequence
the infinitely many KP--hierarchies we noticed in the previous section
are all isomorphic.

At this point the analysis is universal, i.e. it does not depend on
the label $n$,
therefore we simply omit it, and denote $R_n(t_1)$, $S_n(t_1)$
by $R(t_1)$ and $S(t_1)$, respectively.

We construct a bi--Hamiltonian structure for this model as follows.
First we define
\a
Tr(A)=\int\/dx a_{-1}(x),\qquad\qquad
A=\ldots+a_{-1}(x)\partial^{-1}+\ldots\0
\b
for any pseudo--differential operator $A$. Then for any differential
operator $X$ we set $f_X({\hat Q})= <\hat Q X>$. Then we define
\a
\{f_X, f_Y\}_1({\hat Q})&=&<{\hat Q}[Y, X]>\label{PB1}\\
\{f_X, f_Y\}_2({\hat Q})&=&<(X{\hat Q})_+Y{\hat Q}>
-<({\hat Q}_X)_+{\hat Q}_Y>\label{PB2}
\0
\b
The following quantities  are in involution
\a
H_k={1\over k}Tr({\hat Q}^k), \qquad\qquad \forall k\geq1\0
\b
and the two Poisson brackets are compatible
\a
\{H_{r+1},f\}_1=\{H_r, f\}_2\qquad\hbox{\rm for any function}\quad f
\b

On the coordinates $R$ and $S$ the above Poisson brackets are
\a
&&\{R(x), R(y)\}_1=0\qquad\{S(x), S(y)\}_1=0\0\\
&&\{R(x), S(y)\}_1=-\partial_x\delta(x-y)\0
\b
and
\a
&&\{R(x), R(y)\}_2=\bigl(2R(x)\partial_x+R^{'}(x)\bigl)\delta(x-y)
\label{RR2}\\
&&\{S(x), S(y)\}_2=2\partial_x\delta(x-y)\label{SS2}\\
&&\{R(x), S(y)\}_2=\bigl(S(x)\partial_x-\partial^2_x\bigl)\delta(x-y)
\label{RS2}
\b

Remark. For later use we notice that, upon adding the term
\a
+2 \int [ {\hat Q}, Y]_{(-1)}\left(\partial^{-1} [\hat Q, X]\right)\0
\b
to the RHS of(\ref{PB2}), we would obtain
\a
\{S(x),S(y)\}_2=0\label{SS2'}
\b
instead of (\ref{SS2}), while the other two brackets would not change.
However in this case the above compatibility
condition of the two Poisson brackets would not hold as it stands.
It has been shown recently  that such a modified second Poisson
structure appears  in at least two other situations: in the SL(2,R)
WZW model\cite{YW},
and in the conformal affine Liouville theory\cite{AFGZ}.
It is known to be connected with a $W_\infty$ structure.

Let us come now to the flow equations. They can be expressed in compact
form using the two polynomials
\a
F_r(x)={{\delta H_r}\over{\delta S(x)}}
\qquad
G_r(x)={{\delta H_r}\over{\delta R(x)}}
\qquad\forall r\geq1\0
\b
In particular, $F_1=0$ and $G_1=1$.
They satisfy the recursion relations
\a
F^{'}_{r+1}&=&SF^{'}_r-F^{''}_r+2RG^{'}_r+R^{'}G_r\label{F}\\
G^{'}_{r+1}&=&2F^{'}_r+G^{''}_r+\bigl(SG_r\bigl)^{'}\label{G}
\b
and in terms of them
\a
{\partial\over{\partial t_r}}S=G^{'}_{r+1},\qquad\quad
{\partial\over{\partial t_r}}R=F^{'}_{r+1}\label{NLS}
\b
This is the hierarchy characterizing the one-matrix models.
We see immediately that this is the NLS hierarchy. In fact the $t_2$ flow
equations are:
\a
\frac {\partial S}{\partial t_2}= S'' +2S'S + 2R',\quad\quad
\frac {\partial R}{\partial t_2}= -R'' +2(RS)' \label{2flow}
\b
If we change variables as follows
\a
S=(\ln \psi)', \quad\quad\quad R=-\psi\bar\psi\0
\b
and denote by a dot the derivative with respect to $t_2$,
eqs.(\ref{2flow}) become
\a
\dot \psi = \psi''- 2\psi^2\bar \psi, \quad\quad \dot  {\overline\psi}
=-\bar\psi''
+2\bar \psi^2 \psi\0
\b
which contain as a particular case the non--linear
Schr$\ddot {\rm o}$dinger equation. For this reason we will
henceforth refer to this hierarchy as the NLS one.

Next we establish the connection of the matrix
model partition function with
the coordinates of our system as follows. In the discrete case, we have
\a
{\partial\over{\partial t_1}}\ln Z_N(t)=\sum_{i=0}^{N-1}S_i\label{dt1ZN}
\b
Using the first flow
\a
{\partial\over{\partial t_1}}S_n=R_{n+1}-R_n\0
\b
we obtain
\a
{\partial^2\over{\partial t^2_1}}\ln Z_N(t)=R_N\label{R_N}
\b
We will use $R_N$ as the definition of the ``specific heat".
We know that, like any other $R_n$, it satisfies the hierarchy (\ref{NLS})
of equations, so we may simply drop the label N.

Finally, to completely characterize the systems corresponding to
one--matrix models, we need the string equation.
The equation
\a
\relax [\hat Q_n, \hat P_n]=1\label{QP}
\b
does not imply any restriction on the model as long as $P$ has the general
form (\ref{cKP}). One-matrix models are characterized by the following
restriction on the form of $P$
\a
(\hat P_n)_-=0,\quad\quad {\rm i.e.}\quad {\hat P_n}= \sum_{r=1}^\infty rt_r
(\hat Q^{r-1}_n)_+
\label{1mm}
\b
where the label ``+" denotes the pure differential part and
``--" the remaining part of a pseudodifferential operator.
By {\it string equation} we mean (\ref{QP}) together with the
condition (\ref{1mm}). Inserting the flow equations in the string
equation we obtain the Virasoro constraints \cite{BMX}.
\a
L_n Z_N=0, ~~~~~~~n\geq -1\label{Vir}
\b
where
\a
&&L_{-1}=
\sum_{r=2}^\infty rt_r {\partial\over{\partial t_{r-1}}}+Nt_1\0\\
&&L_0= \sum_{r=1}^\infty rt_r {\partial\over{\partial t_{r}}}+N^2
\label{Ln}\\
&&L_n= \sum_{r=1}^\infty rt_r {\partial\over{\partial t_{r+n}}}
+2N \frac{\partial}{\partial t_n} +\sum_{r=1}^{n-1}
\frac {\partial^2} {\partial t_r \partial t_{n-r}},\quad\quad
n>0\0
\b

\section{The KdV hierarchy and the corresponding TFT.}

The KdV hierarchy in the context of
matrix models has been repeatedly analyzed, and our results agree with the
literature. However it is interesting to reconsider the problem from our
point of view, both in itself and as a preparation for the next section.

The KdV hierarchy and the corresponding string equation are obtained
by further restricting the system of the previous section.
This is achieved in the following way. Let us set $S=0$ in (\ref{NLS})
and discard the $t_{2r}$ flows.
Then from eqs.(\ref{F},\ref{G}) we get
\a
&&G_{2r}=0,\label{r1}\\
&&G'_{2r+1}= 2F_{2r}' \label{r2}\\
&&F'_{2r+1}= -F''_{2r}\label{r3}
\b
and
\a
\frac {\partial R}{\partial t_{2r+1}}=
F'_{2r+2}=(\partial^3 +4 R \partial +2R')F_{2r}\label{KdVh}
\b
where the initial condition $F_2=R$ has been used.
Eq.(\ref{KdVh}) is the recursion relation for the KdV hierarchy.
In particular we have
\a
{\partial\over{\partial t_3}}R=R^{'''}
+6RR^{'}\label{KdV}
\b

The connection with the partition function is the same as in the above
section, eq.(\ref{R_N}). This means in particular that the reduced
partition function,  $Z$, is a $\tau$--function of the KdV hierarchy.

It remains for us to find the appropriate form of the Virasoro constraints.
Let us start with the first Virasoro constraint. It was proven in \cite{BX}
that we can put the string equation in the form
\a
\sum_{r=2}^\infty rt_r G'_r+1=0\label{dS}
\b
Using (\ref{r1},\ref{r2}) we find
\a
\sum_{r=1}^\infty (2r+1) t_{2r+1} \frac {\partial R}{\partial t_{2r-1}}
+{1\over 2}=0\label{KdVVir-1}
\b
Essentially from this condition we can obtain the following Virasoro
constraints (see the Appendix for the explicit derivation and for the
problems connected with it)
\a
{\cal L}_n Z=0,~~~~~~~~~~ n\geq -1 \label{KdVVir}
\b
where
\a
&&{\cal L}_{-1}= \sum_{k=1}^\infty (k+{1\over 2})t_{2k+1}
\frac {\partial}{\partial t_{2k-1}} +\frac {t_1^2}{8}\0\\
&&{\cal L}_0=\sum_{k=0}^\infty (k+{1\over 2})t_{2k+1}
\frac {\partial}{\partial t_{2k+1}}+{1\over {16}}\label{VirLn}\\
&&{\cal L}_n=\sum_{k=0}^\infty (k+{1\over 2})t_{2k+1}
\frac {\partial}{\partial t_{2k+2n+1}}+ {1\over 2}
\sum_{k=0}^{n-1} \frac {\partial^2}{\partial t_{2k+1}\partial t_{2n-2k-1}},
{}~~~~n>0\0
\b
Now all the tools we need are at hand.

It has been shown by Witten \cite{W2} and Kontsevitch \cite{Ko} that
the emergence of the KdV hierarchy and the string equation have a
definite topological meaning. In particular, considered in the framework
of a topological field theory coupled to 2D gravity they constitute a very
efficient method to calculate correlation functions.
Following \cite{W2} we define the action corresponding to the KdV system as
\a
S=S_0 - \sum_{k=0}^\infty  t_{2k+1} \int \sigma_{2k+1}^{(2)}\label{action}
\b
The corresponding free energy is
\a
{\cal F}= \sum_{\{n_i\}} {{{ t}_1^{n_1}}\over {n_1!}}{{{t}_2^{n_2}}
\over {n_2!}}\ldots{{{t}_k^{n_k}}\over {n_k!}}
 <\sigma_1^{n_0}\sigma_3^{n_1}\ldots \sigma_{2k+1}^{n_k}>_\circ\label{FE}
\b
where the sum extends over all the n-tuples of integers $\{n_i\}$
=$ \{n_1, n_2, ...,n_k
\}$, and $<\cdot>_\circ$ stands for a correlation function calculated
when all the couplings vanish.
Let us also introduce the notation
\a
{\partial^{n_1}\over{\partial {t}_1^{n_1}}}
{\partial^{n_3}\over{\partial {t}_3^{n_2}}}\ldots
{\partial^{n_k}\over{\partial {t}_{2k+1}^{n_k}}}{\cal F}
\equiv
\ll\sigma_1^{n_1}\sigma_2^{n_2}\ldots \sigma_{2k+1}^{n_k}\gg \0
\b
If we identify this free energy with the one introduced above,
${\cal F}=\ln Z$,
then we have $R={\cal F}''=\ll PP\gg$, where $P\equiv\sigma_1$,  we can use
the KdV hierarchy
and string equation to calculate the correlation
functions for any genus. The calculation is particularly easy
in genus 0 and we will limit ourselves to this case. Then the full KdV
hierarchy can be replaced by the dispersionless KdV hierarchy
\a
{\partial \over {\partial {t}_{2k+1}}}R = 2^k \frac {(2k+1)!!} {k!}
R^kR'\label{dless}
\b
This can be easily obtained by rescaling ${t}_k$ and $R$ by suitable powers
of $N$ and keeping the dominant contribution in ${1\over N}$, which simply
amounts to discarding all the higher derivatives in ${t}_1$ in the flow
equations. A straightforward result is then the following
\a
\ll \sigma_{2k_1+1}\sigma_{2k_2+1}\ldots\sigma_{2k_n+1 }\gg=
\partial^{-2}{\partial^{n} \over {\partial {t}_1^{n}}}
\prod_{i=1}^n\frac {2^{k_i}(2k_i+1)!!}{k_i!}
{{R^{k_1+k_2+...+k_n+1}}\over {k_1+k_2+...+k_n+1}}\label{Sn-2}
\b
where, in the RHS, $\partial^{-1}$ denotes here true integration with respect
to $t_1$ (true integration as opposed to formal integration, since here
we cannot exclude a priori non trivial integration constants).
The expressions of the above correlation functions in terms of the
coupling constants can be found by solving the Landau--Ginzburg
type equation which is obtained by differentiating the ${\cal L}_{-1}$
Virasoro condition
\a
\frac {\partial {\cal F}}{\partial t_{1}}=
2\sum_{k=1}^\infty(2k+1)t_{2k+1}\frac {\partial {\cal F}}{\partial t_{2k-1}}+
\frac {t_1^2}{2}\label{VirL-1}
\b
where we have shifted $t_3\rightarrow t_3-{1\over 6}$. For example the
first critical point is met at $t_{2k+1}=0$ for $k>0$ (small phase space).
In this case we have
\a
<P>= \frac {t_1^2}{2},~~~~~<PP>=R=t_1,~~~~~<PPP>=1\0
\b
$<\cdot>$ denotes a correlation function in the small phase space. Therefore
if we replace $R$ with $t_1$ in (\ref{Sn-2}) we get immediately
the correlation functions $<\sigma_{2k_1+1}\sigma_{2k_2+1}\ldots
\sigma_{2k_n+1 }>$ in the small phase space, up to arbitrary constants
coming from double integration with respect to $t_1$. These integration
constants are determined by the string equation.
For example, from the ${\cal L}_n$ Virasoro condition (\ref{VirLn}),
after extracting
the part relevant to genus 0 and shifting $t_3$ as above, we get
the recursive relation
\a
<\sigma_{n+3}>= 2t_1 <\sigma_{2n+1}> + \sum_{k=0}^{n-1} <\sigma_{2k+1}>
<\sigma_{2n-2k-1}>\0
\b
Similar recursive relations for multiple correlation functions can be
obtained by first differentiating the ${\cal L}_n$ constraints with
respect to the
couplings and then repeating the above derivation.
One can verify that all these recursive relations are satisfied
by
\a
< \sigma_{2k_1+1}\sigma_{2k_2+1}\ldots\sigma_{2k_n+1 }>=
{\partial^{n-2} \over {\partial {t}_1^{n-2}}}
\prod_{i=1}^n\frac {2^{k_i}(2k_i+1)!!}{k_i!}
{{t_1^{k_1+k_2+...+k_n+1}}\over {k_1+k_2+...+k_n+1}}\label{Sn-3}
\b
In the case $n=1$ the first symbol in the RHS has to be understood as
formal integration w.r.t. $t_1$ (i.e. integration with vanishing integration
constant -- we will see in the next section that this is not always the case).
The normalization we used for the $\sigma_{2k+1}$ is completely natural in
the context we presented, but leads to results which differ in normalization
from the ones in the literature. However it is enough to define new
couplings ${\rm t_r}$
\a
{\rm t}_r= 2^{r} \frac {(2r+1)!!}{r!} t_{2r+1},~~~~~r=0,1,2,...\label{relab}
\b
to recover the results of Gross and Migdal \cite{MM}.

Two comments are in order to end this section. What we have done
so far tells us that the correlation functions of the fields $\sigma_k$
can be calculated by using the path integral of one--matrix models
as correlation functions of $Tr (M^{2k+1})$. In other words we have the
correspondence
\a
Tr (M^{2k+1})\leftrightarrow \sigma_{2k+1}\0
\b
The second remark is connected with the first. One should not confuse
the fact that in one--matrix models the conditions $S_n=0$ and
$t_{2k+1}=0$ are related with the fact that we put $S=0$ at the beginning
of this section. What we did is the following: the most general one--matrix
model allowed us to write down an integrable system
(with $S\neq 0$); next we defined a consistent reduction of the system
specified by the condition $S=0$; the reduced theory depends  on the
odd coupling constants. One should bear in mind that the constraint $S=0$
is meant to be
applied to the integrable system, not to the one--matrix model path integral.

\section{The NLS hierarchy and the corresponding TFT.}

In this section we want to do for the larger NLS hierarchy what we have done
in the previous section for the KdV hierarchy; i.e. we try to define a
topological field theory coupled to gravity that corresponds to it.

One quickly realizes that it is not possible to represent this theory simply
by means of the puncture operator $P$ and its descendants as in  the KdV
case. There must be another field, which we call $Q$, coupled through some
parameter in the theory. Since the only parameter in the theory not already
associated to some field is the size $N$ of the lattice, we assume that
$Q$ is coupled to the theory with coupling $N$\footnote{Recently a similar
suggestion was made in ref.\cite{BZ}. We think the spirit of this paper
is close to ours, even though the methods employed are different -- see also
\cite{Gao}.}.
Of course we have to define
the differentiation with respect to $N$. We do it as follows:
for any function
$f_N$ on the lattice we define the derivative $\partial_0$ by means of
\a
\left(e^{\partial_0}-1\right)f_N\equiv f_{N+1}-f_N\label{del0}
\b
Setting $D_0= e^{\partial_0}-1$ we have, conversely,
\a
\partial_0= \ln(D_0+1)\label{D0}
\b
One can easily verify that $\partial_0$ can be identified with
the derivative with respect to $N$ \footnote{It is obvious that when
differentiating with respect to $N$ we understand a continuous extension of
the integer parameter $N$.}.

In particular, using the first flow equations, (\ref{Rn}, \ref{Sn}),
we obtain
\a
&&\left(e^{\partial_0}-1\right)R_N=R_{N+1}-R_N = S_N' \label{dotRN}\\
&&\left(e^{\partial_0}-1\right)S_N=S_{N+1}-S_N=(\ln (R_N+S_N'))'
\label{dotSN}
\b
Next, $R_N$ is connected to the partition function through eq.(\ref{R_N}).
The relation of $S_N$ to the partition function is found by applying
$D_0$ to both sides of eq.(\ref{dt1ZN}). We obtain
\a
\left(e^{\partial_0}-1\right)\frac {\partial}{\partial t_1} \ln Z_N =S_N
\label{S_N}
\b
Applying $D_0$ and ${{\partial}\over{\partial t_1}}$ to
eqs.(\ref{R_N}, \ref{S_N}),
we obtain identities, except when we apply $D_0$ to (\ref{S_N}). In that case
we obtain the following compatibility condition
\a
\left(e^{\partial_0}-1\right)^2 \ln Z_N = \ln (R_N + S_N')+f(N)\label{compat}
\b
where we have integrated once with respect to $t_1$ and $f(N)$ is the
corresponding arbitrary (model dependent) integration constant.

Eqs.(\ref{dotRN},\ref{dotSN},\ref{R_N},\ref{S_N}) and the compatibility
condition (\ref{compat}), together with the NLS flows (\ref{NLS}) and the
Virasoro constraints (\ref{Vir}), are the basis of our  subsequent discussion.
Henceforth, for the sake of uniformity, we relabel the new coupling $N$
as $t_0$. Therefore
\a
t_0=N,\quad\quad\quad\partial_0=\frac{\partial}{\partial t_0}, \qquad
{\rm etc.}\0
\b
Moreover we drop the label $N$ from $R,S$ and $\ln Z$ and understand that
it is included as $t_0$ in the collective label $t$ which represent
the coupling constants.
In particular the first Virasoro constraint  can be rewritten
\a
\sum_{r=2}^\infty rt_r {{\partial {\cal F}}\over{\partial t_{r-1}}}-
\frac {\partial{\cal F}}{\partial t_1} +t_0t_1=0
\label{Vir-1}
\b
where ${\cal F}(t)=\ln Z(t)$ and, for later purposes, we have shifted $t_2$
by $\textstyle -{1\over 2}$ (which sets the first critical point at $t_2=0$).

The result of adding the new coupling $t_0$ can be regarded as an
enlargement of the hierarchy (\ref{NLS}). Beside the $t_k$ flows with
$k\geq 2$ we have now the $t_0$ flow as well. This is obtained by inverting
eqs.(\ref{dotRN}) and (\ref{dotSN})
\a
&&\partial_0 R= S'- {1\over 2} \left( \ln (R+S')\right)''+...\label{0flowR}\\
&&\partial_0 S= \left( \ln (R+S')\right)'- {1\over 2} \left\{\ln \left(
1+ \frac { S' (\ln (R+S'))'}{R+S'}\right)\right\}'+...\label{0flowS}
\b
where dots denote higher order derivatives in $t_1$.

Let us now pass to the field theory language. We want the free energy
${\cal F}$ to be generated by an action
\a
S=S_0 - \sum_{k=0}^\infty t_k \int \sigma_k^{(2)}\label{actionNLS}
\b
The problem is now to express the (perturbed) correlation functions
\a
{\partial^{n_0}\over{\partial t_0^{n_0}}}
{\partial^{n_1}\over{\partial t_1^{n_1}}}\ldots
{\partial^{n_k}\over{\partial t_k^{n_k}}}{\cal F}
\equiv
\ll\sigma_0^{n_0}\sigma_1^{n_1}\ldots \sigma_k^{n_k}\gg \0
\b
in terms of $R$ and $S$. Here
\a
\sigma_0=Q,\qquad\qquad \sigma_1=P\0
\b
So, in particular,
\a
R=\ll PP\gg, \qquad\quad S= \ll (e^Q-1) P\gg \label{defin}
\b
and the compatibility condition (\ref{compat}) means
\a
\ll \left(e^Q-1\right)^2 \gg = \ln \left(\ll PP e^Q \gg\right) +f(t_0)
\label{compat1}
\b

The tools to calculate the correlation
functions are the NLS flows and the string equation, or, equivalently,
the Virasoro constraints (\ref{Vir}) which contain both.
Let us draw first some conclusions concerning the small phase space, i.e.
the space of couplings when all the $t_k$ are set equal
to zero except $t_0$ and $t_1$.
The first Virasoro constraint (\ref{Vir-1}) becomes
\a
<P>=t_0t_1\0
\b
Therefore
\a
<PQ>=t_1,\qquad\quad <PP>=t_0 \label{2point}
\b
and, as a consequence,
\a
<PPQ>=1\label{PPQ}
\b
while all the other correlation functions containing at least one $P$
insertion vanish.
The correlation functions with only $Q$ insertions depend
on the arbitrary function $f(t_0)$ through (\ref{compat1}).

So far the correlation functions look the same as those
of the ${\bf CP}^1$ model studied by Witten, \cite{W1},\cite{DW}.
We will comment later on on this connection. Using the flows (\ref{NLS})
we can write down equations for
\a
\ll \sigma_{n_1}\ldots \sigma_{n_k}Q\gg \qquad {\rm or} \qquad
\ll \sigma_{n_1}\ldots \sigma_{n_k}P\gg \0
\b
in terms of $R,S$ and their derivatives. In order to obtain
\a
\ll \sigma_{n_1}\ldots \sigma_{n_k}\gg \0
\b
we have simply to integrate either the first expression above with respect
to $t_0$ or the second with respect to
$t_1$. The integration constants have to be determined in
such a way as to satisfy the string equation.

Let us consider, as an example, the one point correlation
functions.
The n--th Virasoro constraint can be rewritten,
in the small phase space, as
\a
< \sigma_{n+2}> = t_1 < \sigma_{n+1}> +2t_0 <\sigma_n> +
\sum_{k=1}^{n-1} \left( < \sigma_k \sigma_{n-k}> + < \sigma_k >
< \sigma_{n-k}>\right) \label{Vir-n}
\b
The two--point functions can be obtained by differentiating
the Virasoro constraints with respect to  $t_k$, and so on.
In this way we obtain a full set of
recursive relations that allow us to calculate all the
correlation functions.

The calculation is particularly easy in genus 0 and from now on we limit
ourselves to this case. In order to obtain the equations relevant to genus 0
we rescale
all the quantities by suitable powers of $N$.
\a
{t_k} \rightarrow N^{\frac {2-k}{2}}{t_k} ,
\qquad\quad  R\rightarrow NR, \quad\quad  S\rightarrow N^{1\over 2}S\0
\b
and keep the leading terms in ${1\over N}$.
For example
\a
S= \ll PQ\gg + \sum_{l=1}^\infty {1\over {l!N^{l+1}}} \ll
PQ^l\gg\0
\b
In conclusion in genus 0 we have
\a
R=\ll PP\gg, \qquad\qquad S=\ll PQ\gg \label{0RS}
\b
while the compatibility condition (\ref{compat1}) becomes
\a
\ll QQ \gg = \ln \ll PP \gg +f_0(t_0)\label{QQ0}
\b
where $f_0(t_0)$ is the appropriate genus zero term,
derived from $f(t_0)$ (it is model dependent).
The hierarchy (\ref{NLS}) becomes the dispersionless
hierarchy, i.e.
\a
&&\frac {\partial S}{\partial t_r }= G_{r+1}' =
\sum_{\stackrel{\displaystyle k}{0\leq 2k
\leq r}} \left(\matrix{r\cr 2k\cr }\right) \left(\matrix {2k\cr k\cr}\right)
\left(R^k S^{r-2k}\right)'\label{dlessNLS1}\\
&&\frac {\partial R}{\partial t_r }= F_{r+1}' =
\sum_{\stackrel{\displaystyle k}{2\leq 2k
\leq r+1}} \left(\matrix{r\cr 2k-1\cr }\right) \left(\matrix {2k-1\cr k\cr}
\right) \left(R^k S^{r-2k+1}\right)'\label{dlessNLS2}
\b
Notice that setting $S=0$ and keeping only the odd flows and using
the redefinitions
of the previous section, we obtain (\ref{dless}).

{}From (\ref{dlessNLS1}) we obtain
\a
\ll \sigma_r Q \gg =
\sum_{\stackrel{\displaystyle k}{0\leq 2k
\leq r}} \frac {r!} {(r-2k)! (k!)^2}
\left(R^k S^{r-2k}\right)\label{srQ}
\b
while from (\ref{dlessNLS2}) we get
\a
\ll \sigma_r P \gg =
\sum_{\stackrel{\displaystyle k}{2\leq 2k
\leq r+1}} \frac {r!} {(r-2k+2)! (k-1)!k!}
R^k S^{r-2k+2}\label{srP}
\b

What we said so far for genus 0 is valid in the large phase space.
Now let us come to the the small phase space.
To begin with we have
\a
<QQ>=\ln ~t_0+f_0(t_0)\0
\b
and
\a
R=t_0,\qquad\qquad S=t_1\0\label{RS0}
\b
If we insert this into (\ref{srQ}) and integrate over
$t_0$ (with a vanishing integration constant), we obtain,
in the small phase space
\a
<\sigma_r>=
\sum_{\stackrel{\displaystyle k}{2\leq 2k
\leq r+2}} \frac {r!} {(r-2k+2)! (k-1)!k!}
t_0^k t_1^{r-2k+2}\label{sr}
\b
We can obtain the same result by integrating (\ref{srP})
with respect to $t_1$, but in this case we have to add a suitable
$t_0$--dependent integration constant.

Notice that, both here and in (\ref{srQ},\ref{srP}),
we have made a choice for the integration constants.
This choice can be justified on the basis of the string equation.
For genus 0 eq.(\ref{Vir-n}) takes the form
\a
< \sigma_{n+2}> = t_1 < \sigma_{n+1}> +2t_0 <\sigma_n> + \sum_{k=1}^{n-1}
< \sigma_k > < \sigma_{n-k}>
 \label{Vir-n0}
\b
One can verify that eq.(\ref{sr}) does satisfy
(\ref{Vir-n0}). In a similar way we can derive multi--point
correlation functions.

Let us return to eqs.(\ref{2point}, \ref{PPQ}). If we interpret $P$ and $Q$
as primary fields of an underlying topological conformal field theory, these
equations give us the metric:
\a
\eta_{PQ}=\eta_{QP}=1,\quad\qquad \eta_{QQ}=\eta_{PP}=0\label{metric1}
\b
However a second formulation of the NLS TFT is possible in which $Q$ and $P$
exchange their role, i.e. $Q$ couples to $t_1$ and $P$ to $t_0$. In this case
instead of (\ref{2point}, \ref{PPQ}) we would have
\a
<PQ>=t_1,\qquad\qquad <QQ>=t_0,\qquad\qquad <QQP>=1\label{QQP}
\b
This would give us a diagonal metric
\a
\eta_{PQ}=\eta_{QP}=0,\quad\qquad \eta_{QQ}=1\label{metric2}
\b
while $\eta_{PP}$ is some non--vanishing positive number.
We will refer to this theory as the second formulation of the NLS TFT, while
the previous one will be called the first formulation.
For the second formulation we can of course repeat
the calculations we carried out above for the first.

It is time now to discuss the nature of the NLS topological field theory.
We argue that this field theory belongs to the family of
topological conformal models with two primaries coupled to topological gravity,
\cite{W1}; more precisely it coincides with the explicit version
given in \cite{DW} of  topological ${\bf CP^1}$ sigma model coupled
to gravity, up to some modifications.
We make the comparison only in genus 0, since the latter is explicitly
formulated only in such case. In ref.\cite{DW} one can find recursion relations
and Virasoro constraints in terms of a function $f$. We have proved that,
if $<QQ>= f(<PP>) =\ln <PP>$,
the underlying hierarchy coincides with the dispersionless NLS hierarchy up
to rescaling of the coupling constants. So we can identify the model presented
in \cite{DW} with our first formulation above (with $f_0$ equal to some
constant in (\ref{QQ0})),
provided we choose the above form of the function $f$ (this is not however
the form proposed in \cite{DW}).

On the other hand the model of \cite{DW} coincides with our second formulation,
provided the metric in \cite{DW}
is chosen in such a way that (\ref{QQP}) holds. In this case of course
$<QQ>= \exp <PP>$. The possibility
of these two alternative interpretations was envisaged in \cite{W1}.

\section*{Appendix}

In this Appendix we derive the Virasoro constraints (\ref{KdVVir}).
We start from eq.(\ref{dS}). We can integrate it and obtain
\a
\sum_{k=2}^\infty k t_k G_k+t_1=0\label{KVir-1'}
\b
Actually we should not exclude a priori a non--vanishing integration
constant in the RHS (see below). However this constant must vanish.
One can see it as follows: use the NLS flows and write (\ref{KVir-1'}) as
\a
\sum_{k=2}^\infty k t_k \partial^{-1}\frac {\partial S}{\partial t_{k-1}}
+t_1=0\label{KVir-1''}
\b
Here $\partial^{-1}$ means formal integration (see below).
Next notice that this equation is true for $S=S_n$ for any $n$. Moreover
$S_n = \partial \ln ~h_n$ and $\ln Z_N= \sum _{n =0}^{N-1}\ln~h_n+ {\rm
const}$, see \cite{BMX}. So by summing (\ref{KVir-1'}) over $n$
from 0 to $N-1$,
we obtain the $L_{-1}$ Virasoro constraint (\ref{Vir}) on the lattice.
Viceversa, starting from the latter we can just as easily obtain
(\ref{KVir-1'}).

Now one might be tempted to integrate eq.(\ref{KVir-1'})
once more, but this
does not make sense in the lattice. Therefore we consider (\ref{KVir-1'})
as our starting relation and proceed to the reduction illustrated at the
beginning of section 4. We obtain
\a
\sum _{k=1}^\infty (2k+1)t_{2k+1} \partial^{-1}
\frac {\partial \ln Z}{\partial t_{2k-1}}+t_1=0\label{-1'}
\b
We now integrate this equation with respect to $t_1$ and obtain
\a
{\cal L}_{-1}Z=b_{-1}Z\label{-1}
\b
which is almost the first Virasoro condition except for the integration
constant $b_{-1}$, where $b_{-1}\equiv b_{-1}(t_3,t_5,\ldots)$.

Next we apply the operator
\a
{\cal D}= \partial^3 +4R\partial +2R'\0
\b
to eq.(\ref{-1'}). Using (\ref{KdVh}) and integrating twice, we obtain
\a
{\cal L}_{0}Z=(a_0t_1+b_{0})Z\label{0}
\b
where $a_0$ and $b_0$  are integration constants depending on
$t_3,t_5,\cdots$.
Notice that we have included a part of the integration constant
$={1\over 16}$ in the definition of ${\cal L}_0$.
We could continue in this way by integrating once (\ref{-1}) and applying
${\cal D}$ and so on. We would obtain a series of integration constants
$a_n$ and $b_n,~~n>0$ similar to the above ones. In order to understand
how to calculate them we have to clarify preliminarily a few points.

When considering $R$ and polynomials
of $R$ and its derivatives with respect to $t_1$ we are entitled to use
the identities $\partial \partial^{-1}=1$ and $\partial^{-1}\partial =1$.
In this sense we speak of the symbol $\partial^{-1}$ as a formal integration.
The reason is that $R$ and polynomials of $R$ and its derivatives are
supposed to belong to a family of objects to which pseudodifferential
calculus applies (for example, to the family of functions  rapidly
decreasing at infinity in the variable $t_1$ ).
However this is in general not guaranteed for an infinite sum such as, for
example, $\sum_{k=2}^\infty k t_k G_k$. If we insist on applying
$\partial^{-1}$ to
such objects as a formal integration, we are bound to run
into inconsistencies. Therefore when applying $\partial^{-1}$  to the
infinite sums that appear in the string equations we have to interpret
it as a true integration and take care of the corresponding integration
constants.

On the other hand the KdV hierarchy and its solutions are
characterized by homogeneity in $t_n$ with degree assignment
\a
deg(t_n)=n\0
\b
We have consequently
\a
deg({\cal L}_n)=-2n = deg(a_nt_1+b_n)\label{deg}
\b
This is the first ingredient we will use. The second is the form of the
constant $a_n$ and $b_n$. A rather general form we can imagine for them
is a sum of expressions like
\a
\sim~\prod_{i=1}^{s} t_{2k_i+1}^{a_i}\label{form}
\b
where $a_i$ are real numbers. We can strongly restrict the enormous
number of possibilities implied by equation (\ref{form}) by using the
path integral form of the model we are studying. We remember that in the
path integral the terms involving $t_n,~~n\geq 3$ can be considered as
perturbations. Therefore we do not expect anything dramatic to happen
when only one of these couplings is set to zero. Thus we can conclude
that in (\ref{form}) all the exponents $a_i$ are non--negative
\footnote{This is certainly what happens in the
lattice, see (\ref{Vir}).}. If we now make a degree analysis
of the various possible constants, we conclude that only
$b_{-1}$ and $b_0$ can be non--vanishing, and the latter is a true
constant, i.e. it does not depend on $t_3,t_5,...$.
What remains for us to do is to calculate these two constants.
To this end we can repeat the analysis of the Appendix of \cite{BMX}.

Let us summarize the situation. So far we have found the relations
\a
{\cal L}_{n}Z=b_{n}Z\label{-n}
\b
where only $b_{-1}$ and $b_0$ are non--vanishing. Since
\a
\relax [{\cal L}_n,{\cal L}_m]=(n-m){\cal L}_{n+m}\0
\b
the consistency conditions
\a
\relax [{\cal L}_n, b_m]-[{\cal L}_m, b_n] = (n-m)b_{n+m}
\label{consis}
\b
must be satisfied. Take these equations for $m=1$ and $m=0$. Studying
the $t_1$ dependence we find
\a
\frac {\partial b_{-1}}{\partial t_{2k+1}}=0,~~~~~~~~~~~~~
\frac {\partial b_{0}}{\partial t_{2k+1}}=0, ~~~k\geq 0\0
\b
So $b_{-1}$ and $b_0$ are true constants (for $b_0$ this was already known
independently to us). If we use (\ref{consis}) for $n=0$ and $m=-1$
and for $n=1$ and $m=-1$ we immediately conclude that
\a
b_{-1}=0, \qquad\qquad b_0=0\0
\b
respectively. This complete our proof of (\ref{KdVVir}).

A few comments. Our proof relies on the form (\ref{form}) for the
integration constants and on path integral considerations. This means
that we cannot completely exclude the existence of other reductions of
the NLS hierarchy to the KdV one (beside the one of section 4),
in which the Virasoro conditions (\ref{KdVVir}) are replaced
by more complicated conditions (due to non--vanishing integration
constants). However we are rather skeptical about this possibility since
in the lattice case it was proved without any assumption, \cite{BMX},
that only in the $L_{-1}$ and $L_0$ Virasoro conditions do non
trivial integration constants appear -- and are of a very simple form.

\vskip 0.5cm
\noindent
{\bf Acknowledgements}
\vskip 0.5cm

We would like to thank L.L.Faddeev and J.Schiff for interesting
discussions. One of us (L.B.) would like to thank the Instituto
de Fisica Teorica -- UNESP for the kind hospitality extended to him
during the completion of this paper and FAPESP for partial financial
support.


\begin{thebibliography}{}

\bibitem{MM} E. Brezin and V. Kazakov, Phys.Lett.{\bf B236}(90)144;

             M. Douglas and S. Shenker, Nucl.Phys.{\bf B335}(90)635;

             D. Gross and A. Migdal, Phys.Rev.Lett.{\bf 64}(90)127;

             T. Banks, M. Douglas, N. Seiberg and S. Shenker,
Phys.Lett.{\bf B238}(90) 279;

             M. Douglas, Phys.Lett.{\bf B238} (90) 176.

\bibitem{MMR} V.A.Kazakov, in Proc. Carg\'ese workshop in 2d gravity,
O.Alvarez et al. eds. O.Alvarez, E.Marinari and P.Windey;

              L.Alvarez-Gaum\'e, Helv.Phys.Acta{\bf 64}(1991) 361;

              P.Ginsparg, lectures at 1991 Trieste Summer School,
Los Alamos preprint.


\bibitem{W1}  E.Witten, Nucl.Phys.{\bf B340} (1990) 281.

\bibitem{W2}  E.Witten, Surveys in Diff.Geom.{\bf 1}(1991) 243.

\bibitem{DW} R.Dijkgraaf and E.Witten, Nucl.Phys.{\bf B342}(1990) 486.

\bibitem{Ko} M.Kontsevich, Comm.Math.Phys.{\bf 147}(1992) 1.

\bibitem{D} R.Dijkgraaf, {\it Intersection theory, integrable hierarchies
and topological filed theory}, IASSNS--HEP--1/91.

\bibitem{BX} L.Bonora and C.S.Xiong, Phys.Lett.{\bf B285}(1992) 191.

\bibitem{AFGZ} H.Aratyn, L.A.Ferreira, J.F.Gomez and A.H.Zimerman,
{\it On two--current realization of KP hierarchy}, IFT--P/020/92

\bibitem{YW} F.Yu and Y.--S. Wu, Phys.Rev.Lett.{\bf 68}
(1992) 2996.

\bibitem{BMX} L.Bonora, M.Martellini and C.S.Xiong,
Nucl.Phys.{\bf B375}(1992)453.

\bibitem{BIZ} D.Bessis, C.Itzykson and J.--B.Zuber, Adv.Appl.Math. 1 (1980)
109.

\bibitem{DVV} R. Dijkgraaf, H. Verlinde, and E. Verlinde,
Nucl.Phys.{\bf B348}(1991)435.

\bibitem{FKN} M.Fukuma, H.Kawai and R.Nakayama,
Int.J.Mod.Phys.{\bf A6} (1991) 1385.

\bibitem{BZ} E.Br\'ezin and J.Zinn--Justin, Phys.Lett. B288 (1992) 54.

\bibitem{Gao} H.B.Gao, {\it On renormalization group flow in matrix models}
ICTP preprint IC--302/92.



\end{thebibliography}
\end{document}